\documentclass[twocolumn,showpacs,preprintnumbers,amsmath,amssymb,nofootinbib]{revtex4}

\usepackage{graphicx}
\usepackage{dcolumn}
\usepackage{bm}
\usepackage{amsfonts}
\usepackage{verbatim}

\newcommand{\vnabla}{{\mbox{\boldmath$\nabla$}}}
\newcommand{\veta}{{\mbox{\boldmath$\eta$}}}

\newcommand{\vA}{{\mbox{\boldmath$A$}}}
\newcommand{\vB}{{\mbox{\boldmath$B$}}}

\newcommand{\vD}{\mbox{\boldmath$D$}}

\begin{document}


\title{Influence of the domain walls on the Josephson effect in Sr$_2$RuO$_4$}

\author{A. Bouhon}
\email{abouhon@phys.ethz.ch}
 
\author{M. Sigrist}
 
\affiliation{Institute for Theoretical Physics, ETH-Zurich, Zurich
  Switzerland}%

\date{\today}%

\begin{abstract}
A detailed theoretical interpretation of the Josephson interference experiment between Sr$_2$RuO$_4$ and Pb reported by Kidwingira \textit{et al} \cite{Maeno2006} is given. Assuming chiral p-wave pairing symmetry a Ginzburg-Landau theory is derived in order to investigate the structure of domain walls between chiral domains. It turns out that anisotropy effects of the Fermi surface and the orientation of the domain walls are essential for their internal structure. Introducing a simple model for a Josephson junction the effect of domain walls intersecting the interface between Sr$_2$RuO$_4$ and Pb is discussed. It is shown that  characteristic 
deviations of the Fraunhofer interference pattern for the critical Josephson current as a function of the magnetic field occurs in qualitative agreement with the experimental finding. Moreover the model is able also to account for peculiar hysteresis effects observed in the experiment. 
\end{abstract}

      
\maketitle

\section{Introduction}

The aim to identify the symmetry of the superconducting phase in Sr$_2$RuO$_4$ has stimulated numerous experiments since more than a decade \cite{REVMOD,PHYSTODAY}. Most of them provide strong evidence that the pairing state has the so-called chiral $p$-wave symmetry. This state is the analog of the A-phase of superfluid $^3$He, which is a spin triplet state breaking  time reversal symmetry with an angular moment along the fourfold $ c $-axis of the tetragonal crystal lattice of Sr$_2$RuO$_4$. The full gap function is a $2 \times 2 $-matrix in spin space, $ \hat{\Delta}_{\vec{k}} $, and can be represented by the $ \bm{d} $-vector in the case of spin-triplet pairing:
\begin{equation}
\bm{d} (\bm{k}) = \Delta_0 \hat{\bm{z}} (k_x \pm i k_y) = - tr \left\{ \hat{\Delta}_{\vec{k}}  i \hat{\sigma}_y  \hat{\bm{\sigma}} \right\}   \; ,
\end{equation}
with $ \Delta_0 $ as the gap magnitude.
SQUID interferometer experiments probing the internal phase structure of the Cooper pairs are consistent with odd-parity ($p$-wave) pairing \cite{LIU}.
NMR Knight shift measurements are compatible with
equal-spin spin triplet pairing with the spin axis lying in the basal plane of the tetragonal crystal lattice, i.e. $ \bm{d} \parallel \hat{\bm{z}} $ \cite{ISHIDA}. Muon spin relaxation studies show enhanced internal magnetism in the superconducting phase suggesting a state with broken time reversal symmetry \cite{LUKE}. Similarly, recent Kerr effect studies imply the presence of an orbital magnetic moment pointing along the $c$-axis \cite{xia:167002}. On the other hand, the search for the spontaneous magnetization at the surface of samples,  as expected for a chiral $p$-wave state, by
highly sensitive scanning probes has only given negative results so far \cite{tamegai,moler}.

A further extraordinary property of the chiral $p$-wave state is its two-fold degeneracy which can lead to domains, which are distinguished by
the orientation of the orbital angular momentum of the Cooper pairs, parallel or antiparallel to the c-axis. 
Early on the question of the formation of domains of the two states (angular momentum up and down) has been discussed, but no experimental indications of domains had been reported, until recently. Kidwingira et al investigated carefully the interference pattern of a Josephson junction between Sr$_2$RuO$_4$ and the conventional superconductor Pb in a magnetic field \cite{Maeno2006}. They
interpreted these patterns as the result of domain walls intersecting the extended Josephson junction
and so giving rise to a spatial variation of the Josephson phase along the junction. It has been speculated in the past that domain walls intersecting the interface between a chiral $p$-wave and conventional superconductor could influence the Josephson effect, as the Josephson current-phase relation could be different for the two types of domains \cite{asano}. Such a property could lead to an intrinsically inhomogeneous junction and would alter the interference pattern. This situation bears some similarity with the
$45^{\circ} $ asymmetric interface in oriented films of high-$T_c$ superconductors where
faceting of the boundary introduces a random switch of Josephson phase by $ \pm \pi $ along the interface \cite{MANNHART}. This type of  boundaries display unusual interference patterns of the critical current in a magnetic field and generate spontaneous flux pattern on the interface. Unlike in the standard case the maximum of the critical current is here usually not located at zero field.

This finding could eventually provide indirect evidence for the presence of domain walls in the $p$-wave superconductor. The discussion of this problem requires a detailed analysis of the domain wall structure. For interpretation of their results Kidwingira et al introduced several types of domain walls which could give rise to strong variations of the Josephson phase \cite{Maeno2006}. While they are possible on a topological level, not all the proposed domain walls are energetically stable. Indeed in the most simple approach assuming full rotation symmetry for the Fermi surface
around the $z$-axis of the $p$-wave superconductor none of the stable domain walls give rise to any shift of the Josephson phase and the presence of domain walls could go basically unnoticed
in the interference experiment \cite{asano}. However, as we will show below a more careful analysis taking more general conditions of the electronic spectrum into account leads to stable and metastable domain walls which can give rise to non-trivial intrinsic phase patterns in a Josephson junction and reproduce the experimentally observed anomalies in the interference pattern.

\section{ \label{GLT} Ginzburg-Landau free energy}

\begin{figure}[htbp]
\begin{center}
\includegraphics[scale=0.8]{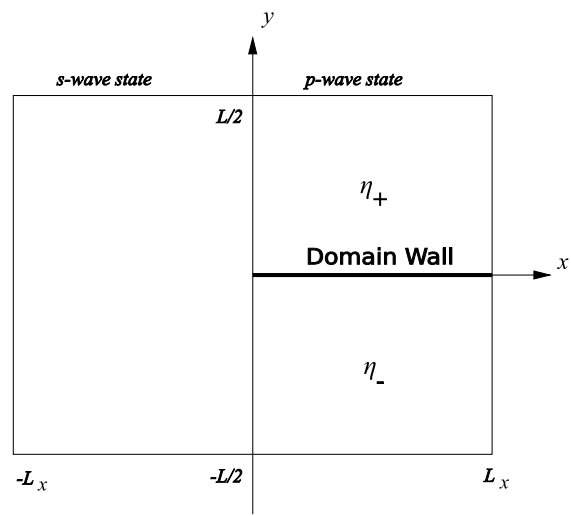}
\end{center}
\caption{\small Josephson Junction between the s-wave state and the chiral p-wave state with a domain wall parallel to the crystal $x$-axis between the two degenerate states: $ \veta_{\pm}\propto(\pm 1,i)$. The dimensions of the junction are $\Delta x\times \Delta y\times \Delta z=2L_x\times L \times L_z$.}
\label{JJ2}
\end{figure}

Our study on the Josephson effect involves both a conventional $s$-wave and a chiral $p$-wave superconductor, representing the experimental arrangement of Pb coupled to Sr$_2$RuO$_4$ \cite{Maeno2006}. We first introduce here the basic order parameters and their
corresponding Ginzburg-Landau theories which will be used later to discuss the structure of domain walls in the chiral $p$-wave state and their influence on the Josephson effect in configurations, as shown in Fig. \ref{JJ2}. 

The conventional superconductor is described by a scalar order parameter $ \psi(\bm{r}) $ for the spin-singlet pairing state of highest possible symmetry (''$s$-wave" pairing state). The Ginzburg-Landau free energy functional has the standard form,
\begin{equation}
\mathcal{F}_s[\psi,\vA]
        =\int_{V_s} d^3r \biggr[ a_s(T) \vert \psi\vert^2+b_s\vert\psi\vert^4+K_s\vert\vD\psi\vert^2 + \frac{\vB^2}{8\pi}\biggl]  \;,
        \label{FS}
\end{equation}
with $\vD=\vnabla-i\gamma\vA$  as the gauge-invariant derivative and $\gamma=2e/\hbar c=2\pi/\Phi_0$ ($\bm A $ is the vector potential with the magnetic field $ \bm{B} = \bm{\nabla} \times \bm{A} $), \ $\Phi_0$ is the magnetic flux quantum, $a_s (T) = a_s' (T-T_{cs}) $, $ b_s $ and $K_s$ are parameters \cite{Gennes1999}.

The chiral $p$-wave phase requires a two-component order parameter $ \bm{\eta} = (\eta_x , \eta_y) $ with
\begin{equation}
\bm{d} (\bm{k}) = \hat{\bm{z}} (\eta_x k_x + \eta_y k_y ) \; ,
\end{equation}
which belongs to the irreducible representation $ E_u $ of the tetragonal point group $ D_{4h} $ of Sr$_2$RuO$_4$. The free energy functional has then the following general form
\begin{equation} 
\begin{array}{l}
 \mathcal{F}_p[\eta_x,\eta_y,\vA] =
\displaystyle{\int_{V_p}} d^3r \biggr[ a_p(T) |\bm{\eta}|^2 + b_1 | \bm{\eta}|^4  \nonumber\\
\\
  \qquad + \dfrac{b_2}{2} ({{\eta_x}^*}^2 {\eta_y}^2 + {\eta_x}^2 {{\eta_y}^*}^2)  + b_3 \vert \eta_x \vert^2 \vert \eta_y \vert^2  \nonumber\\
  \\
  \qquad  + K_1 \bigr(\vert D_x \eta_x \vert^2 + \vert D_y \eta_y \vert^2\bigl)
        + K_2 \bigr(\vert D_x \eta_y \vert^2 + \vert D_y \eta_x \vert^2\bigl) \nonumber\\
        \\
  \qquad + \bigr\{  K_3 (D_x\eta_x)^* (D_y\eta_y) + K_4 (D_x\eta_y)^* (D_y\eta_x)+c.c. \bigl\}   \nonumber\\
  \\
  \qquad +  K_5 (\vert D_z \eta_x \vert^2 + \vert D_z \eta_y \vert^2) + \dfrac{\vB^2}{8\pi} \biggl]  \;.
        \label{FP}
        \end{array}
\end{equation}
The coefficients $b_i$ and $K_i$ are material-dependent parameters and $ a_p = a_p' (T-T_{cp}) $ \cite{Sigrist1991}.
In order to stabilize the chiral $p$-wave state in the bulk the coefficients have to satisfy the
relations: $ b_2 > 0 $, $ b_2 > b_3 $ and $ 4b_1 > b_2 - b_3 $. Minimizing the free energy functional with respect to $\vert\eta_0\vert$ in the homogeneous case, we obtain the uniform phase\footnote{We note that the corresponding bulk energy density is given by $f_{homog}[\eta_0]=(-4b_1+b_2-b_3)\eta_0^4=a_p\eta_0^2=-H_c^2/8\pi$, which defines the critical magnetic field $H_c$.} 
\begin{equation}
\bm{\eta}_{\pm}  = \eta_0 ( 1 , \pm i )\; ,  \quad |\eta_0| ^2 = \frac{-a_p(T)}{4 b_1 -b_2 + b_3} \;,
\label{eta0}
\end{equation}
such that the corresponding gap function is
\begin{equation}
 \bm{d}_{\pm} (\bm{k}) = \eta_0(T) \hat{\bm{z}} (k_x \pm i k_y)  \;.
 \end{equation}
The two states with opposite relative sign are degenerate and violate time reversal symmetry, as the operation of time reversal $ \hat{K} $ yields $ \hat{K} \bm{d}_{\pm} = \bm{d}_{\pm}^* = \bm{d}_{\mp} $.
These two states can form domains. In the following we will first
analyze the structure of domain walls  between such domains of opposite chirality and  their influence on the Josephson effect in geometries as shown in Fig.\ref{JJ2}.

\section{Domain wall structure}

Before addressing the influence of the domain wall on the Josephson effect it is necessary to analyze here the structure of a domain wall in detail. Our aim is to show which types of
domain walls are energetically favorable. It will turn out that the
anisotropy of the Fermi surface plays an important role and with this also the orientation of the domain wall. 
For a convenient discussion it will be advantageous to change the order parameter representation of the $p$-wave superconductor and to formulate the Ginzburg-Landau functional also in rotated coordinate frames. 

\subsection{Free energy functional formulation}

In a first state we express now the free energy functional in a new form which simplifies the rotation of the coordinate frame. We then always define the domain wall as the $z$-$x$-plane and the spatial variation of the order parameter occurs only along the corresponding $y$-axis perpendicular to this plane. For simplicity we keep the $z$-axis as fixed along the crystaline $z$-axis.\\ 

We introduce the parametrization \cite{Sigrist1999},
\begin{equation}
\eta_\pm=\frac{1}{2}(\pm\eta_x-i\eta_y) \;, \quad D_\pm=D_x\pm iD_y \; ,
\end{equation}
which inserted into the free energy leads to
\begin{equation} 
\begin{array}{l}
\mathcal{F}[\eta_+,\eta_-,\vA] =
\displaystyle \int_{V_p} d^3r \biggr[ 2a_p (|\eta_+|^2+|\eta_-|^2) \nonumber \\
\\
 \qquad+ b\bigr\{ |\eta_+|^4+|\eta_-|^4  + 4\vert\eta_+\vert^2\vert\eta_-\vert^2   \nonumber \\
 \\
 \qquad + \nu ({{\eta_+}^*}^2 {\eta_-}^2 +{\eta_+}^2 {{\eta_-}^*}^2)\bigl\}  \nonumber \\
 \\
 \qquad+ K \Bigr\{ \vert \vD \eta_+ \vert^2 + \vert \vD \eta_- \vert^2
                -\dfrac{1}{2} \bigr( \nu (D_- \eta_+)^* (D_+ \eta_-) \nonumber \\
                \\
 \qquad + (D_+ \eta_+)^* (D_- \eta_-)+c.c.\bigl) \Bigl\}  + \dfrac{1}{8\pi} (\vnabla \times \vA)^2\biggl]  \; .
 \end{array}
\label{Fpm}
\end{equation}
In a weak-coupling approach the parameter $\nu=(\langle v_x^4\rangle-3\langle v_x^2v_y^2\rangle)/(\langle v_x^4\rangle+\langle v_x^2v_y^2\rangle)$ is a measure for the anisotropy of the Fermi surface ($\nu=0$ for a cylindrical Fermi surface and $\vert\nu\vert=1$ for a square-shaped Fermi surface) where $v_{x,y,z}$ are the components of the Fermi velocity and $\langle \cdot\rangle$ defines the average over the Fermi surface.  The coefficients in Eq.(\ref{FP}) satisfy the following relations
\begin{eqnarray}
\begin{array}{ccc}
 b_1&=&b\frac{3+\nu}{8} \; ,\\
 b_2&=&b\frac{1-\nu}{4}  \; , \\
 b_3&=&-b\frac{1+3\nu}{4}  \; ,
\end{array}
\begin{array}{ccc}
K_1&=&K\frac{3+\nu}{4}  \; ,\\
K_2=K_3=K_4&=&K\frac{1-\nu}{4}  \; . \\
 & &
\label{bK}
\end{array} 
\end{eqnarray}
where $ b$ and $ K $ are material dependent parameters.

This form of the free energy functional allows us now to deal easily with the rotation of the reference frame around the $z$-axis: $(x',y')=(x\cos\theta+y\sin\theta,y\cos\theta-x\sin\theta)$ with the angle $ \theta $ relative to the original $x$-axis in the tetragonal crystal. For the
new coordinate frame, the order parameter and the gradients are transformed as $\eta_\pm=\mathrm{e}^{\mp i\theta}\eta_\pm'$ and $D_\pm=\mathrm{e}^{\pm i\theta}D_\pm'$. When we express  (\ref{Fpm}) in the new coordinates, we have only to modify the two terms containing the parameter $\nu$ by phase factors,
\begin{eqnarray*}
&&b \nu ~ \mathrm{e}^{i4\theta}{{\eta_+'}^*}^2 {\eta_-'}^2+c.c.~~~~,\\
&&-\frac{K}{2}(\nu \mathrm{e}^{i4\theta}(D_-' \eta_+')^* (D_+' \eta_-')
+(D_+' \eta_+')^* (D_-' \eta_-')+c.c.)  .
\end{eqnarray*}
In this way we can use the angle $ \theta $ to define the coordinate frame.
In the following we will omit the primes and always assume that we describe the domain wall in the corresponding frame. 

\subsection{Variational ansatz for the domain wall structure}

We now turn to the structure of the domain wall which will depend qualitatively on the choice of parameters in the free energy. We choose here a variational approach to discuss 
behavior of the order parameter around the domain wall with the following ansatz which is most useful to eventually obtain the key information relevant for the Josephson effect,
\begin{equation}
\label{apan}
 \eta_+=\eta_0\mathrm{e}^{i\phi_+}\cos\chi~~\mathrm{and}~~\eta_-=\eta_0\mathrm{e}^{i\phi_-}\sin\chi,
\end{equation}
with the boundary conditions within a given reference frame,
\begin{eqnarray}
\label{BC}
\chi=\left\{
\begin{array}{cc}
 0 &~~~ y \to +\infty\\
\frac{\pi}{2} &~~~ y\to-\infty
\end{array}\right. \;,
\end{eqnarray}
where we restrict the spatial dependence of the order parameter to the function $\chi (y)$ and use the relative phase $\alpha=\phi_+ - \phi_-$ as a further (variational) parameter. In this way the domain wall appears as an interface between the two superconducting domains which is additionally characterized by a phase difference $ \alpha $, similar to a Josephson junction or weak link \cite{Sigrist1999}. 

For the later discussion of the Josephson effect with an $s$-wave superconductor, it will be useful to return to the order parameter components $ (\eta_x , \eta_y ) $ which are expressed  as
\begin{eqnarray}
 \begin{array}{ccc}
\eta_x&=&\eta_0(\mathrm{e}^{i\phi_+}\cos\chi-\mathrm{e}^{i\phi_-}\sin\chi)  \;,\\
\eta_y&=&i\eta_0(\mathrm{e}^{i\phi_+}\cos\chi+\mathrm{e}^{i\phi_-}\sin\chi)  \;.
\end{array}
\end{eqnarray}
The chosen boundary conditions in (\ref{BC}) correspond to the situation,
\begin{eqnarray}
\begin{array}{l}
\veta (y=-\infty) = \eta_0(-1,i) \mathrm{e}^{i\phi_-}   \equiv \veta_-   \;,\\
\veta (y=+\infty) = \eta_0(+1,i) \mathrm{e}^{i\phi_+} \equiv \veta_+ \;,
\end{array}
\end{eqnarray}
for which, in the parametrization $ (\eta_x,\eta_y)=(\vert \eta_x\vert\mathrm{e}^{i\phi_x},\vert \eta_y\vert\mathrm{e}^{i\phi_y})$, the phase shifts of the order parameters are given by 
\begin{eqnarray}
\label{DeltaPhi}
\begin{array}{ccccc}
\Delta\phi_x&=&\phi_x(+\infty)-\phi_x(-\infty)&=&\alpha-\pi  \;,\\
\Delta\phi_y&=&\phi_y(+\infty)-\phi_y(-\infty)&=&\alpha  \;.
\end{array}
\end{eqnarray}
Here $\alpha$ plays the role of the total phase difference of the order parameter and also determines the current flow through the domain wall analogous to a Josephson junction. 

With this variational ansatz we rewrite the free energy, 
\begin{equation}
\begin{array}{l}
\mathcal{F} = \displaystyle \int d^3r \biggr[ -b\eta_0^4+\dfrac{b\eta_0^4}{2}(1+\nu\cos(2\alpha-4\theta))\sin^2(2\chi)\nonumber \\
\\
   \qquad + \dfrac{(\vnabla \times \vA)^2}{8\pi} + K\eta_0^2  \Bigr\{  \vert \vD \cos\chi\vert^2 + \vert \vD \sin\chi\vert^2 \nonumber \\
   \\
   \qquad   - \frac{1}{2} \bigr( \mathrm{e}^{-i\alpha} (D_+ \cos\chi)^* (D_-\sin\chi) \nonumber \\
   \\
   \qquad+ \nu\mathrm{e}^{i(\alpha-4\theta)} (D_+ \sin\chi)^* (D_-\cos\chi)+c.c. \bigl) \Bigl\}  \biggl] \;, \nonumber \\
\end{array}
\end{equation}
where we substitute $a_p=-b\vert \eta_0\vert^2$ using (\ref{eta0}) and (\ref{bK}). Assuming homogeneity along the $z$- and $x$-axis, we take $\chi=\chi(y)$ and $ {\bm A} = {\bm A}(y) $ with  $A_z=0$,  the free energy can be written as
\begin{equation}
\begin{array}{l}
 \mathcal{F} = \displaystyle \int d^3r \biggr[ -b\eta_0^4+ K\eta_0^2  \biggr\{  \dfrac{Q}{4\xi_0^2}\sin^2 2\chi + (\partial_y\chi)^2 \nonumber \\
 \\
     \qquad + \gamma^2 (A_x^2+A_y^2) + C_+ \sin2\chi \left(\gamma^2(A_y^2-A_x^2)-(\partial_y\chi)^2\right) \nonumber \\
     \\
     \qquad+ 2 S_- \gamma^2 A_x A_y \sin 2\chi  + 2 \gamma \partial_y\chi \left(S_+ A_y-C_- A_x\right)\biggl\} \nonumber \\
     \\
     \qquad+ \dfrac{(\partial_yA_x)^2}{8\pi}  \biggl]  \;,
\end{array}
\label{FAxy}
\end{equation}
where  $ \xi_0^2=K/2b\eta_0^2$ defines the coherence length. Additionally we introduced
\begin{equation}
\begin{array}{l}
 Q      = 1+\nu\cos(2\alpha-4\theta), \\
C_{\pm} = (\cos\alpha\pm\nu\cos(\alpha-4\theta))/2,  \\
S_{\pm} = (\sin\alpha\pm\nu\sin(\alpha-4\theta))/2  \; .
\end{array}
\end{equation}
which are the only coefficients depending on the anisotropy parameter $ \nu $,  the phase $ \alpha $ and the angle $ \theta $ of the domain wall orientation relative to the crystalline main axis (crystal $x$-axis). 

Certain symmetries become immediately obvious here.
The free energy is invariant under the rotation $\theta\rightarrow\theta\pm\pi/2$ and the operation $(\theta,\nu)\rightarrow(\theta\pm\pi/4,-\nu)$ leaves the coefficients unchanged. It is also clear that the phase $ \alpha $ is closely linked to the orientation of the domain wall. A further aspect of symmetry is connected with the operation $ (\theta , \alpha) \rightarrow (-\theta, -\alpha) $, leading to $ Q \rightarrow + Q $ , $ C_{\pm} \rightarrow + C_{\pm} $ and $ S_{\pm} \rightarrow - S_{\pm} $. 

The variational minimization of the free energy functional with respect to $\chi$, $A_x$ and $A_y$ leads to the corresponding three
equations:
\begin{equation}
\begin{array}{l}
\partial_y^2\chi = \dfrac{Q}{4\xi_0^2} \sin4\chi-\gamma (S_+ \partial_y A_y-C_- \partial_y A_x)+ C_+ \sin2\chi \partial_y^2\chi \nonumber\\
\\
   \qquad +\cos2\chi \Bigr[C_+ \bigr( \gamma^2 (A_y^2-A_x^2) - (\partial_y\chi)^2\bigl) + 2S_- \gamma^2 A_xA_y \Bigl]\;,
    \nonumber\\
\\
  \kappa^2\xi_0^2\gamma ~\partial_y^2 A_x - \gamma (1-C_+ \sin 2\chi)  A_x = S_- \gamma A_y\sin 2\chi \nonumber\\
\\
  \qquad - C_- \partial_y\chi \;, \nonumber\\
\\
  A_y = -\dfrac{S_+  \partial_y\chi + S_- \gamma A_x\sin 2\chi }{\gamma(1+C_+ \sin 2\chi)}\;.
\end{array}
\label{min-equ}
\end{equation}
For a concise notation we introduce the Ginzburg-Landau parameter $\kappa^2=(\lambda/\xi_0)^2=1/(8\pi K \eta_0^2 \gamma^2 \xi_0^2)$ with $ \lambda=[1/(8\pi \gamma^2 K \eta_0^2)]^{1/2}$ as the London penetration length (we take $\hbar = 1$). The third equation leads to $A_y\rightarrow -A_y$ under $ (\theta,\alpha)\rightarrow (-\theta,-\alpha)$. As a consequence we observe that the free energy is invariant under the
operation $  (\theta,\alpha,A_y)\rightarrow (-\theta,-\alpha,-A_y) $, which will be important again later in the analysis of the Josephson effect. 

These equations will now be solved numerically, although we will also present in \ref{AA} an approximate analytical solution.
For this purpose it is useful to turn to the dimensionless variables, measuring lengths in units of $ \xi_0 $, $\tilde{y}=y/\xi_0$, and
using for the vector potential  $a_i=\gamma \xi_0 A_i$. The free energy per unit area of the domain wall can then be written as
\begin{equation}
\begin{array}{l}
 f   =  \dfrac{K \eta_0^2}{\xi_0}  \displaystyle \int d\tilde{y} \biggr[ -\frac{1}{2}
         + \dfrac{Q}{4} \sin^2 2\chi + (\partial_{\tilde{y}}\chi)^2 + a_x^2 + a_y^2 \nonumber\\
         \\
         \qquad + C_+ \sin2\chi \left(a_y^2-a_x^2-(\partial_{\tilde{y}}\chi)^2\right) + 2 S_- a_x a_y \sin2\chi \nonumber \\
         \\
     \qquad + 2 \partial_{\tilde{y}}\chi \left(S_+ a_y-C_- a_x\right) + \kappa^2 (\partial_{\tilde{y}}a_x)^2 \biggl]  \;.
 \end{array}
  \label{fdw}
\end{equation}
The numerical results for the spatial dependence for the order parameter and the vector potential derived from $\chi$, $a_y$ and $a_x$ minimizing  $f$ are shown in \ref{sol_eta_phi}.

\begin{figure}[htbp]
\begin{center}
\includegraphics[scale=1.3]{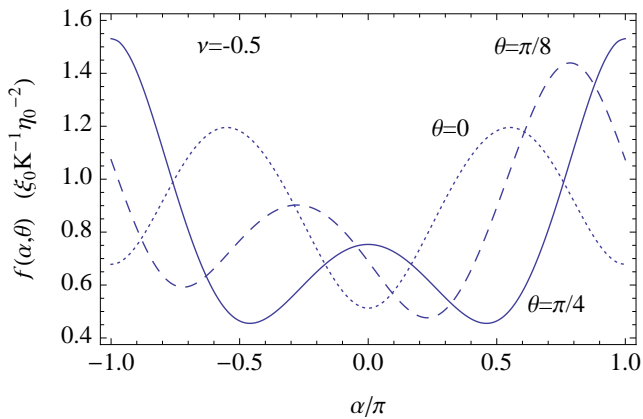}
\end{center}
\caption{\small Domain wall energy as a function of the phase difference $\alpha=\phi_+-\phi_-$ for rotations of the domain wall by the angles $\theta=0$ (solid line), $\pi/8$ (dashed line) and $\pi/4$ (dotted line). The anisotropy parameter is chosen $\nu=-0.5$.}
\label{sola}
\end{figure}

\begin{figure}[htbp]
\begin{center}
\includegraphics[scale=1.3]{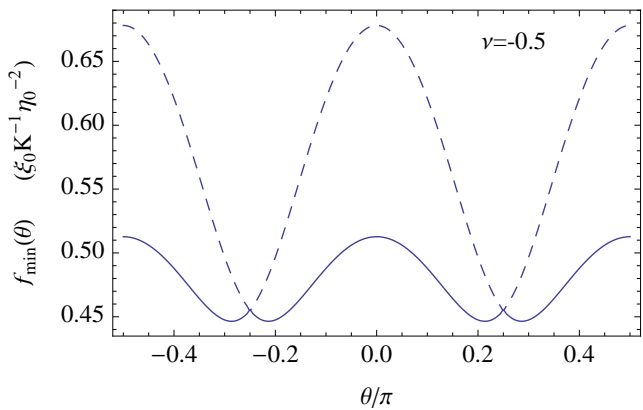}
\end{center}
\caption{\small Energy of stable (solid line) and metastable states (dashed line) of the domain wall obtained through the minimization with respect to $\alpha$: $f_{min}(\theta)=f(\alpha_{min},\theta)$, as a function of the angle of the domain wall and for $\nu=-0.5$.}
\label{fmin_nu05}
\end{figure}

\begin{figure}[htbp]
\begin{center}
\includegraphics[scale=1.3]{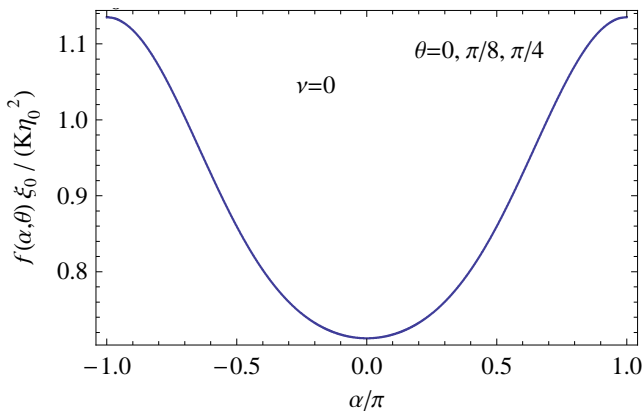}
\end{center}
\caption{\small Domain wall energy for an isotropic Fermi surface, i.e. $\nu=0$.}
\label{sol0}
\end{figure}

In Fig. \ref{sola} we show the domain wall energy per unit area as a function of $ \alpha $ and $ \theta $. We do not use here $ \alpha $ as a variational parameter but search for the variational local minimum for given $ \alpha $. For this calculations we choose the anisotropy parameter to be negative, $ \nu = -0.5 $. Note that the result for $ \nu = 0.5 $ follows immediately from the symmetry relation, $ f(\alpha, \theta,\nu) = f(\alpha, \theta \pm \pi/4, - \nu) $. 
For $ \theta =\pi/4 $ we observe two degenerate minima of $ f(\alpha, \theta,\nu) $ at $ \alpha_{min} \approx \pm 0. 46 \pi $ corresponding to the stable domain wall configuration for given $ \theta $ and $ \nu $. For any other angle $ \theta $ this degeneracy is lifted leading to a stable and metastable minimum, which are located at $ \alpha= 0 $ and $ \alpha = \pi $, respectively, if $ \theta = 0 $. 

The energy of the stable and metastable domain wall configuration depends on the angle $ \theta $, as can be seen in Fig. \ref{fmin_nu05} where we plot  $f_{min}(\theta) \equiv f(\alpha_{min}(\theta),\theta)$ for $\nu=-0.5$ (the solid line marks the stable and the dashed line the metastable states). We find that a minimum of this energy occurs at an angle $ \theta \approx \pm 0.21\pi $ away from $ \theta = 0 $. Analogously by symmetry such a minimum is found at $ \theta \approx \pm( \pi/4-0.21\pi) = \pm 0.04 \pi $ for $ \nu = 0.5 $. From this we conclude that there are special orientations for the domain wall which are energetically favorable and depend on the anisotropy properties of the superconductor. These special  orientations need not to lie along symmetry axes or symmetry planes. Note that also the angle with respect to the $z$-axis is important in this respect, since domain walls parallel to $x$-$y$-plane are probably most stable. However, here we consider only the case of domain walls 
parallel to the $z$-axis, as they are generally most important for the modification of the Josephson effect. 

As a reference we consider also the case of an isotropic Fermi surface ($\nu=0$)  which naturally does not show any dependence on the angle $ \theta $. We find that the most stable domain wall state corresponds to the phase $ \alpha =0 $ which agrees with the result obtained from a corresponding microscopic model calculation based on a quasi-classical approach  \cite{Sigrist1999b}. 

\begin{figure}[htbp]
\begin{center}
\includegraphics[scale=1.3]{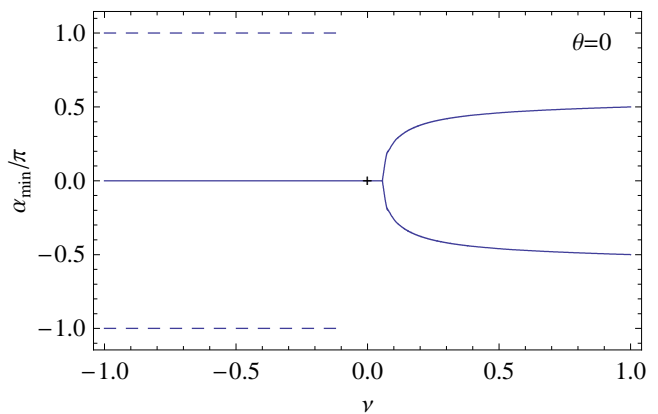}
\end{center}
\caption{\small The stable and metastable values $\alpha_{min}$ as a function of the anisotropy of the Fermi surface $\nu$, for a domain wall parallel to the crystal $x$-axis, i.e. $\theta=0$. The solid lines correspond to stable states while the dashed line denotes metastable values $ \alpha_{metastable} = \pm \pi $.}
\label{Dtheta0}
\end{figure}

In Fig. \ref{Dtheta0} we show the behavior of $ \alpha_{min} $ as a function of $ \nu $ for the angle $ \theta =0 $. There are two obvious regions; for $ \nu < \nu_c = 0.057 $ the minimum corresponds to $ \alpha_{min} =0 $ and for $ \nu > \nu_c $ we find two degenerate values. In addition metastable states (indicated as dashed lines) appear at $ \alpha_{metastable} = \pm \pi $ for $ \nu < -0.12$. 

\subsection{Domain wall at the surface}

In view of our later discussion of the Josephson junctions intersected by domain walls we now consider the situation of a domain wall ending at the surface of the superconductor and being pinned at a defect somewhere in the bulk (for illustration we simplify our model introducing a columnar pinning center along the $z$-axis). 
We now search for the possible domain wall configurations. The domain wall has to compromise between being as short as possible and as close as possible to the orientation (angle $\theta$) minimizing its wall energy whereby, in principle, both stable and metastable (local minimum) situation can play a role. 

We restrict ourselves here to the case of $ \nu <  0 $ which is relevant for Sr$_2$RuO$_4$ \cite{Agterberg1998,Agterberg1999}. 
Let us start with the situation that the surface normal vector $ {\bm n} $  is assumed to be along a crystal main axis in basal plane, say the $x$-axis.

\begin{figure}[htbp]
\begin{center}
\includegraphics[scale=0.8]{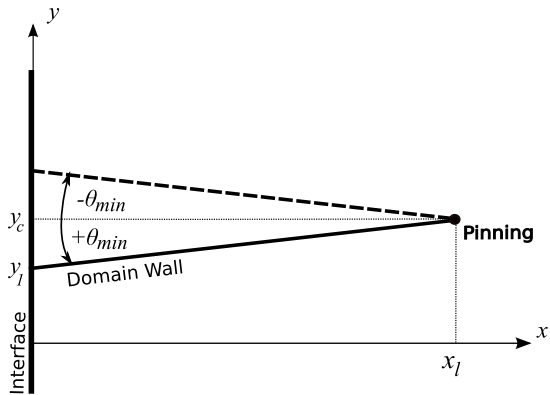}
\end{center}
\caption{\small Two possible configurations of the domain wall: the domain wall may jump  between the two positions if it is driven by an external magnetic field, as shown below.}
\label{DWjump}
\end{figure}

\begin{figure}[htbp]
\begin{center}
\includegraphics[scale=1.3]{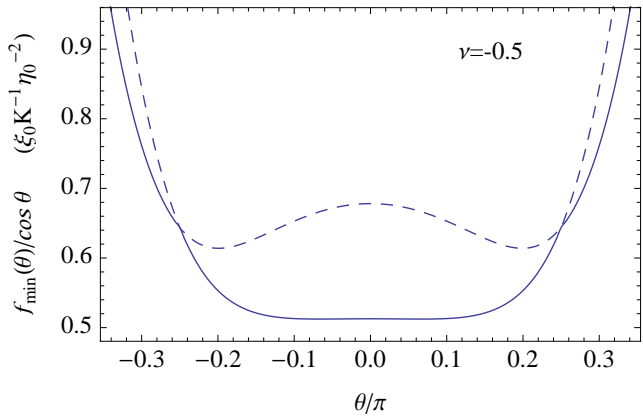}
\end{center}
\caption{\small Stable branch (solid line) and metastable branch (dashed line) of the minimum energy for a domain wall at the surface : $f_{min}(\theta)/\cos\theta$, for $\nu=-0.5$.}
\label{fminc_nu05}
\end{figure}

The columnar defect along the $z$-axis is located at a distance $ x_l $ from the surface (Fig. \ref{DWjump}). The domain wall energy per unit length in $z$-direction for a phase $ \alpha $ and an angle $ \theta $ relative to the $x$-axis is given by
\begin{equation}
E_d(\alpha, \theta) = x_l \frac{f(\alpha , \theta)}{\cos \theta} \; .
\end{equation}
with $ f(\alpha, \theta) $ defined in Eq.(\ref{fdw}).
This is a symmetric function under $ (\alpha, \theta) \to  (-\alpha, -\theta) $ and plotting $ E_d (\alpha,\theta) $ we find two stable situations with $ (\alpha_{min}, \theta_{min} ) $ and
$ (-\alpha_{min}, -\theta_{min}) $. The two configurations are depicted in Fig.\ref{DWjump}.  We show in Fig. \ref{fminc_nu05} the corresponding domain wall energies of the stable and metastable branch given by the solid line and the dashed line respectively. The lowest energy contributions correspond to the two rather shallow minima of the stable branch. 

\begin{figure}[htbp]
\begin{center}
\includegraphics[scale=1.3]{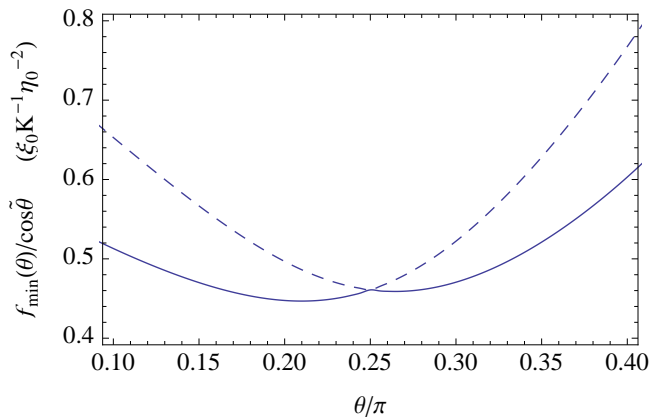}
\end{center}
\caption{\small Stable branch (solid line) and metastable branch (dashed line) of the minimum energy for a domain wall at the surface for $\nu=-0.5$, when the surface normal vector makes an angle $\bar{\theta}=0.2\pi$ with the crystal $x$-axis ($\tilde{\theta}=\theta-\bar{\theta}$).}
\label{fmin_nu05m_bar02}
\end{figure}

If the surface normal vector does not lie along a main axis (but within the $x$-$y$-plane) then the situation becomes more complex. While in the previous case two degenerate stable
domain wall states were found, under general conditions only two local minima of the domain wall energy exist which are not degenerate. 
In Fig. \ref{fmin_nu05m_bar02} we show the domain wall energy for an angle  $ \bar{\theta} =0.2\pi$ of the normal vector $ {\bm n} $ relative to the $ x$-axis. Because the domain walls for given orientation possesses phases $ \alpha $  for stable and metastable energy minima, there are
two branches of Fig. \ref{fmin_nu05m_bar02} which correspond to possible domain wall configuration at the surface. These configurations are neither degenerate nor symmetric unlike for the case of $ {\bm n } =(100) $. Nevertheless, they represent domain wall states which are, in principle, accessible depending on the history of the system as there are two local energy minima in Fig. \ref{fmin_nu05m_bar02}, a stable (global minimum) and a metastable (local minimum).

\section{Josephson coupling between $s$- and $p$-wave superconductors}

The symmetry aspects of the  Josephson coupling between a conventional $s$-wave and an odd-parity superconductor have been discussed many years ago \cite{Sigrist1991,Geshk1986}. Considering the standard lowest-order Josephson coupling we are confronted with the apparent two-fold obstacle that there is a mismatch in the spin configuration (singlet versus triplet) as well as in parity of the orbital part of the two pair wave functions. It has been shown, however, that spin-orbit coupling and the reduced inversion symmetry at the interface
are sufficient to yield a finite coupling. The lowest-order Josephson coupling for an interface with normal vector $ \bm{n} $ and the $s$-wave order parameter $ \psi $ and the $p$-wave order parameter $ \bm{\eta} $ has then the form \cite{Sigrist1991,Geshk1986}, 
\begin{equation}
\label{junctionenergy}
\mathcal{F}_J = -t(\bm{n})  \int_{i} dS \left\{ \psi^* (\bm{\eta} \times \bm{n}) \cdot \hat{\bm{z}}  + \psi   (\bm{\eta}^* \times \bm{n}) \cdot \hat{\bm{z}} \right\} \; ,
\end{equation}
if the $ \bm{d} $-vector is parallel to $z$-axis as for the chiral $p$-wave state. Here $ t(\bm{n}) $ denotes the coupling strength and the integral runs over the interface. Unlike the coupling between two singlet-superconductors, which can be estimated through the experimentally determined normal state tunneling conductance, in our case the difference in the pseudo-spinors yields spin-active (spin-flip) tunneling processes which depend on the spin-orbit coupling of the two materials. For both Pb and Sr$_2$RuO$_4$ spin-orbit coupling is not small, such that it can be expected that the Josephson coupling is comparable to an ordinary, although a reliable estimate is not easy due to the complex band structure.

Using $ \mathcal{F}_J $ as a boundary term in the Ginzburg-Landau equations we can derive the following Josephson current-phase relation of the junction:
\begin{equation} 
 \label{JCG}
\begin{array}{ll}
J &\displaystyle = \frac{2e K_s}{i \hbar} \left\{ \psi^* \bm{n} \cdot \bm{\nabla} \psi - c.c. \right\} \\
 & = \displaystyle \frac{2e t}{i \hbar} \left[ \psi^* (\bm{\eta} \times \bm{n}) \cdot \hat{\bm{z}} - c.c. \right] \;.
 \end{array}
 \end{equation}
We assume $ \bm{n} = (1,0,0) $ such that the coupling reduces to
 \begin{equation}
 J = - \frac{2et}{i \hbar} \left\{ \psi^* \eta_y - \psi \eta_y^* \right\} = +\frac{4 e t}{\hbar} | \psi | | \eta_y| \sin \varphi = J_0 \sin \varphi \;,
 \end{equation}
with $ \varphi = \phi_y - \phi_s $ as the phase difference between the order parameters of the p-wave and the s-wave superconductor. In this geometry only the $ \eta_y $-component of the $ p$-wave side order parameter contributes to the Josephson coupling. This is the order parameter component 
of the pairing state whose nodes point towards the interface. Only this component can combine with the spin to conserve the total angular momentum of 
the Cooper pair in the tunneling through the interface.

\section{Interference pattern}

We want to calculate the Josephson critical current assuming that several domain walls intersect the Josephson junction from the Sr$_2$RuO$_4$ side. 
In the following we model the effect of the domain wall as a step-like phase shift at the point of intersection. This approach is justified, if we consider the
Josephson coupling as weak such that the currents are small and the Josephson penetration depth $ \Lambda_J $ defined below is longer than the extension 
of the junction and even much longer than the width of the domain walls \cite{Bouhon2007}.

\subsection{Junction in a uniform magnetic field}

Analogous to ordinary Josephson junctions we find also in junctions between an $s$- and a chiral $p$-wave superconductor a relation between the derivative of the phase $ \varphi = \phi_y - \phi_s $ with respect to the coordinate along the contact and the magnetic field threading perpendicularly. Thus with the assumption that  the junction normal vector is $ 
\bm{n} = (1,0,0) $, we examine the variation of the phase $ \varphi $ and magnetic field $B_z $ along the 
$ y$-direction and obtain, 
\begin{equation}
\partial_y \varphi (y) = \frac{2 \pi}{\Phi_0} d_{eff} B_z(y) + \partial_y \phi(y) \; ,
\label{phase-grad}
\end{equation}
where the effective width of the Josephson contact (parallel to $ \bm{n} $) $ d_{eff} = d + \lambda_s + \lambda_p $ includes the real width $ d $ and the London penetration depths on both sides, $ \lambda_{s,p} $ (for Pb $ \lambda_s \approx 35 nm $ and for Sr$_2$RuO$_4$ $ \lambda_p \approx 160 nm $).  
In contrast to the standard case we add here a contribution $ \partial_y \phi $ which corresponds to the intrinsic phase variation on the $ p$-side which, for example, can be induced by a domain wall intersecting the Josephson contact, as we will discuss below. However, also faceting of the interface can
introduce such contributions. 

The spatial variation of the phase $ \varphi(y) $ obeys the extended Sine-Gordon equation,
\begin{equation}
\partial_y^2 \varphi = \frac{1}{\Lambda_J^2} \sin \varphi + \partial_y^2 \phi    \;,
\end{equation}
with Josephson penetration depth $ \Lambda_J  = \{ c \Phi_0 / 8 \pi^2 J_0 d_{eff} \}^{1/2} $. Assuming now that $ \Lambda_J $ is larger than the extension $ L $ of the Josephson junction along the $y$-direction, we find that, for a uniform external field $H$,  $ \varphi $ is given approximately by
\begin{equation}
\partial_y \varphi = \frac{2 \pi}{\Phi_0} d_{eff} H + \partial_y \phi \quad \Rightarrow \quad \varphi(y) = k y + \phi(y) + \beta  \;,
\end{equation}
where $ k = 2 \pi H d_{eff} / \Phi_0 $ and $ \beta $ is an integration constant. The total current is then
\begin{equation}
I = L_z J_0 \int_{-L/2}^{+L/2} dy \sin (ky + \phi(y) + \beta)  \;,
\label{current-1}
\end{equation}
with $ L_z $ the extension of the junction along the $z$-direction. This allows us now to discuss the effect of domain wall on the interference pattern in the maximal Josephson current obtained by maximizing $ I $ with respect to $\beta$.

\subsection{Phase $ \phi(y)$}

Before addressing the interference pattern it is necessary to determine the change of the phase $ \phi(y) $ as we pass through a domain wall along the interface of the Josephson junction. This phase enters  Eq. (\ref{phase-grad}) through the relation
\begin{equation}
\partial_y \phi_y - \frac{2 \pi}{\Phi_0} A_y =  \partial_y \phi \; ,
\end{equation}
on the side of the $p$-wave superconductor. Consider now a single domain wall which reaches the junction interface at a $90^{\circ}$-angle, i.e. $ \theta = 0 $. The phase shift $ \Delta \phi $ 
through the domain wall can be determined by integrating along the $ y$-axis,
\begin{eqnarray} 
\Delta \phi (\theta =0 )  &=&  \int_{- \infty}^{+\infty} dy \; \partial_y \phi_y - \frac{2 \pi}{\Phi_0}  \int_{- \infty}^{+\infty} dy A_y\nonumber\\
							&=&	 \pm \alpha (\theta=0) - \frac{2 \pi}{\Phi_0}  \int_{- \infty}^{+\infty} dy \; A_y, 
\end{eqnarray}
where $ A_y $  has been calculated in Eq.(\ref{min-equ}). The plus (minus) sign of the phase $ \alpha $ reflects the two possible situations for the domain wall: moving along the $y$-axis we cross a domain wall from $ \eta_-  \to \eta_+ $ ($+$-sign)  or vice versa ($-$-sign). Since the two situations are related through the transformation $ y \to - y $ it is clear that the 
phase shift $\Delta\phi$ is identical in absolute magnitude for the two cases. 

In general, this phase difference depends also on the angle of intersection $ \tilde{\theta} $, the angle between interface normal vector $ {\bm n} $ (in $x$-$y$-plane) and domain wall. A rotation by an angle $ \tilde{\theta} $ involves a phase shift $ e^{\pm i \tilde{\theta}} $ for $ \eta_\pm $. First we consider the case $\eta_- \to \eta_+ $ for which $\Delta\phi_y = \phi_+ -\phi_-=\alpha$, if $ {\bm n} $ is parallel to the domain wall. Rotating $ {\bm n} $ by the angle $ \tilde{\theta}$, we obtain $\Delta\phi'_y =  \phi'_+ - \phi'_- = (\phi_+ + \tilde{\theta}) - (\phi_- - \tilde{\theta})=\alpha+2 \tilde{\theta}=\alpha'$. Thus the phase shift through the domain wall is then given by
\begin{eqnarray}
\label{deltaphi_definition}
\Delta \phi(\tilde{\theta}) &=& \alpha(\tilde{\theta}) - \frac{2 \pi}{\Phi_0}  \int_{- \infty}^{+\infty} dy \; A_y \; \nonumber\\
				&=& \alpha'(\tilde{\theta}) - 2 \tilde{\theta} \nonumber\\
						  &-& \frac{2\pi}{\Phi_0}\displaystyle \int_{-\infty}^{+\infty} (\cos^2 \tilde{\theta} A_y' + \cos\tilde{\theta} \sin\tilde{\theta} A_x') dy' ,
\end{eqnarray}
where $A_y'$ and $A_x'$ are solutions of Eq.(\ref{min-equ}) for $\tilde{\theta}\neq0$. 

For an interface with $ \bm{n} = (100) $ the most stable states are characterized by the two orientations of the domain wall $\pm\theta_{min}$, according to the previous discussion.  By symmetry, we know that the free energy is invariant under the transformation $(\alpha,\theta,A_y)\rightarrow (-\alpha,-\theta,-A_y)$. Changing from  $\theta_{min}$ to $-\theta_{min}$ and keeping the same boundary conditions ($+\alpha_{min}(-\theta_{min})=-\alpha_{min}(\theta_{min})$), the phase shift simply changes sign
\begin{equation}
\Delta\phi (-\theta_{min})= -\Delta\phi(\theta_{min})  \;.
\end{equation}
Similarly, going from the configuration $\veta(\pm\infty)=\veta_{\pm}$ to  $\veta(\pm\infty)=\veta_{\mp}$ (keeping the same angle $\theta_{min}$), i.e. under a time reversal transformation, only the sign of the  phase shift is changed 
\begin{equation}
\Delta\phi_+(\theta_{min})  \longrightarrow \Delta\phi_-(\theta_{min}) = -\Delta\phi_+(\theta_{min})  \;,
\end{equation}
where we have introduced the notation $\Delta\phi_+\equiv\Delta\phi(\veta_- \to \veta_+)$ and $ \Delta\phi_-\equiv\Delta\phi(\veta_+ \to \veta_-)$. We note that this is in agreement with the fact that the later transformation is equivalent to an inversion of the boundary conditions done by the transformation $\alpha\rightarrow -\alpha$, which causes a sign change for the phase difference. 

Therefore, to each domain wall intersecting the Josephson junction we can associate a phase difference 
$ \Delta \phi_+ (\theta_{min}) $ (resp. $ \Delta \phi_- (\theta_{min}) $) corresponding to the geometry $\veta(y=\pm\infty)= \veta_{\pm}$ (resp. $\veta(y=\pm\infty)=\veta_{\mp}$) which depends on the orientation of the domain wall. 

For orientation of the interface normal vector different from $ \bm{n} = (100) $, we compare the two lowest energy configurations of the domain wall (stable and metastable) state.  Using the above calculation scheme we can show that the phase differences are in general different for the two states, yielding $ \Delta \phi_{1} $ and $ \Delta \phi_{2} $. Under time reversal operation, changing between the two cases $ {\bm \eta}_+ \to {\bm \eta}_- $ and $ {\bm \eta}_- \to {\bm \eta}_+ $, we obtain the opposite sign of the phase shifts again. Note that, in general, the two phase shifts, 
$ \Delta \phi_{1} $ and $ \Delta \phi_{2} $ are comparable in magnitude with those cases where $ {\bm n} $ lies along a symmetry axis.  Note, however, that there is an energy difference between the two domain wall states, if they are not degenerate. This energy difference depends on the 
depth over which the domain wall is influenced by the change of configuration at the interface.

\subsection{Model for intersecting domain walls}

We extend now our model to a Josephson junction which is intersected by several domain walls. First consider the case of one domain wall reaching the interface at the position $ y = y_1 $, which yields the Josephson phase, 
\begin{equation}
\varphi(y) = ky + \phi(y) + \beta = ky+\Delta\phi_{1 \mu} \Theta(y-y_1)+\beta \; .
\end{equation}
Here $ \Delta \phi_{1 \mu} $ denotes the phase shift at the domain wall as calculated in the previous section. The index $ \mu $ labels the two types of domain wall state (stable or metastable minima). 
It is admissible to use a step function $\Theta(y)$ to describe the spatial change of the phase, since the extension of the domain wall is small compared to the length of the Josephson junction and the Josephson penetration depth $ \Lambda_J $. This leads to a piecewise constant phase shift. 

The generalization to an array of domain walls is straightforward. For $N$ successive domain walls, the Josephson phase difference across the junction is given by
\begin{equation}
\varphi(y) = \beta+\sum_{i=1}^{N}\Delta\phi_{i \mu_i} ~\Theta(y-y_{i})+ky \;,
\label{model-1}
\end{equation}
where $ \varphi(y) $ depends on the configuration $ \{ \mu_1, \mu_2, \dots ,\mu_N \} $ of all domain walls. We will now show that the distortion of the Josephson interference pattern through $ \varphi(y) $
depends on these configurations. 

\subsection{Modified interference pattern}

The total Josephson current which traverses the junction is given by Eq.(\ref{current-1}). Using our  piece-wise constant approximation for $ \phi(y) $ we obtain readily the following expression, 
\begin{eqnarray}
\frac{I(\Phi)}{I_0}=\displaystyle \frac{\Phi_0}{\pi\Phi}
\sum_{i=0}^{N}\sin\left(\beta+\sum_{j=0}^{i}\Delta\phi_{j \mu_{j}} +\frac{\pi\Phi}{\Phi_0}\frac{(y_{i+1}+y_{i})}{L}\right)\nonumber\\
\times \sin\left(\frac{\pi\Phi}{\Phi_0}\frac{(y_{i+1}-y_{i})}{L}\right) ,~~~
\label{model-2}
\end{eqnarray}
with $ \Delta \phi_{0 \mu_0} = 0 $, $y_{N+1}=L/2$, $y_{0}=-L/2$ and $ I_0=S J_0$ ($ S=L_z L$ being the interface area), and $ k L \Phi_0  = 2\pi\Phi $. 

The Josephson critical current, i.e. the maximal supercurrent possible for this configuration, is obtained by the maximization of $ I$ with respect to $\beta$. This condition reads
\begin{eqnarray}
\frac{\partial I}{\partial \beta}(\beta_{max})=0\;,
\end{eqnarray}
leading to
\begin{eqnarray}
\beta_{max}=\arctan\left(\frac{A_1}{A_2}\right)\;,
\end{eqnarray}
with
\begin{eqnarray}
 \qquad \displaystyle A_1 = \sum_{i=0}^{N}\cos\left(\sum_{j=0}^i \Delta\phi_{j \mu_j} +\frac{\pi\Phi}{\Phi_0}\frac{(y_{i+1}+y_{i})}{L}\right) \nonumber\\
 \times \sin\left(\frac{\pi\Phi}{\Phi_0}\frac{(y_{i+1}-y_{i})}{L}\right)\;,\nonumber \\
\nonumber \\
\qquad \displaystyle A_2 =\sum_{i=0}^{N}\sin\left(\sum_{j=0}^i \Delta\phi_{j \mu_j} +\frac{\pi\Phi}{\Phi_0}\frac{(y_{i+1}+y_{i})}{L}\right) \nonumber\\
\times \sin\left(\frac{\pi\Phi}{\Phi_0}\frac{(y_{i+1}-y_{i})}{L}\right)\;.
\end{eqnarray}
The Josephson current can be written as
\begin{equation}
\frac{I}{I_0}=\frac{\Phi_0}{\pi\Phi} \{ A_1\sin\beta +A_2\cos\beta \} \;,
\end{equation}
and finally we find
\begin{eqnarray}
\frac{I_{max}}{I_0} &=& \left\vert\frac{\Phi_0}{\pi\Phi}\right\vert ~(A_1\sin\beta_{max}+A_2\cos\beta_{max}) \nonumber\\
&=& \left\vert\frac{\Phi_0}{\pi\Phi}\right\vert ~\sqrt{A_1^2+A_2^2} \;.
\label{model-3}
\end{eqnarray}

\begin{figure}[htbp]
\begin{center}
\includegraphics[scale=1.1]{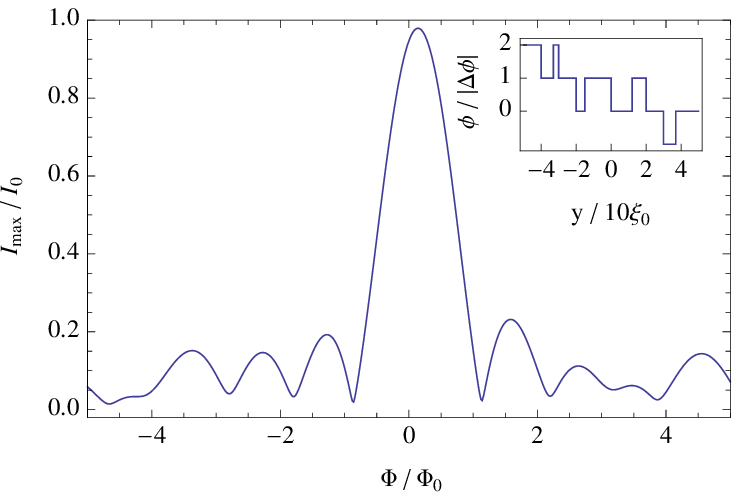}
\end{center}
\caption{\small Josephson critical current as a function of the external magnetic flux for ten intersecting domain walls randomly configured with a total phase shift $2\vert\Delta\phi\vert$. The anisotropy parameter is chosen $\nu=-0.6$ \cite{Agterberg1998,Agterberg1999}.}
\label{Imax_06m}
\end{figure}

\begin{figure}[htbp]
\begin{center}
\includegraphics[scale=1.1]{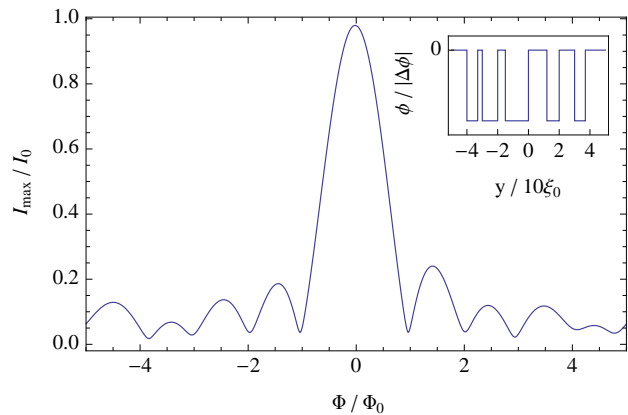}
\end{center}
\caption{\small Josephson critical current as a function of the external magnetic flux for ten intersecting domain walls randomly configured with a zero total phase shift. The anisotropy parameter is chosen $\nu=-0.6$ \cite{Agterberg1998,Agterberg1999}.}
\label{Imax_06m2}
\end{figure}

In Fig.\ref{Imax_06m} and Fig.\ref{Imax_06m2} we show two simulations for interference pattern of the Josephson critical current in a magnetic field, using two random domain wall configurations (indicated in the inserted panels) where we assumed $ N =10 $ for the number of intersecting domain walls.  The randomness occurs in the positions of domain walls as well as in the sequence of $ \{ \mu_1, \dots , \mu_N \} $. The deviation from the standard Fraunhofer pattern is obvious. However, the maximum of the Josephson current lies rather close to $ H = 0 $. This feature has to do with the fact that, 
in the case of generic random configurations, the phase shift does not vary much overall, i.e. $ | \phi(y=L/2) - \phi(y=-L/2) | \ll 2 \pi $. Stronger deviations can be observed, if we bias the domain wall configuration in a way as to have a larger net shift, e.g. by assuming for all domain walls the same sign of $ \Delta \phi $.

\subsection{Hysteresis and noise effects}

An important feature supporting the idea that domain walls are involved in producing the irregular interference pattern is the observation of hysteresis effects by Kidwingira \textit{et al},  when the external magnetic field was cycled between positive and negative maximal fields \cite{Maeno2006}. They argued that the applied magnetic field induces a rearrangement of the domain walls in the sample, and they substantiated their claim by simulations looking at the effect of shifted domain wall positions. Since the domain walls are pinned at defects of the sample large shifts in positions are rather unlikely.  

Hysteresis effects are rather easily discussed within our simplified model in Eq.(\ref{model-1}, \ref{model-2}, \ref{model-3}). The free energy of the junction in a magnetic field can be approximated by
\begin{eqnarray}
&F&(\beta, \Phi, \{\mu_1, \dots , \mu_N \} )\nonumber\\
&=& \frac{I_0 \Phi_0}{2 \pi c L }  \sum_{i=0}^N \int_{y_i}^{y_{i+1}} dy \cos \left(\beta + \sum_{j=0}^i \Delta \phi_{j, \mu_j} + y \frac{\pi \Phi}{\Phi_0}  \right)\; . \nonumber\\
\label{fen}
\end{eqnarray}
We neglect the change of the intersection points $ y_i $ for different configurations $ \mu_i $ and examine the condition to minimize the free energy. Moreover we assume that the interface is smooth and that we have the same two essentially degenerate configurations for all domain walls:
$ \Delta \phi_{i 1} = \Delta \phi_+ > 0 $ and $ \Delta \phi_{i 2} = \Delta \phi_- < 0 $. Thus, $ {\bm n} $ is assumed to lie very close to a high-symmetry axis of Sr$_2$RuO$_4$, say $ {\bm n} = (100) $.
For values $ \Phi > 0 $ the free energy in Eq.(\ref{fen}) can be lowered by choosing $ \Delta \phi_{i \mu_i} = \Delta \phi_- $, because under this condition
\begin{equation}
\sum_{j=0}^i \Delta \phi_{j, \mu_j} \approx \Delta \phi_- \frac{N(y+L/2)}{L} < 0  \; .
\label{polar-phi}
\end{equation} 
Note that there are special values of $ \Phi $ where the full ''polarization'' of $ \Delta \phi $ may not be the best choice. However, considering a sweep of the field to maximal value $ \Phi_{max} $ and back would favor Eq.(\ref{polar-phi}). The same argument can be used for negative fluxes, driving the domain walls to adopt $ \Delta \phi_{i \mu_i} = \Delta \phi_+ $. Note that  for metastable domain wall states a ''polarization''  would only be possible, if the energy expense of the metastable configurations is sufficiently small. We consider now geometries for which this is true. 
  
\begin{figure}[htbp]
\begin{center}
\includegraphics[scale=1.1]{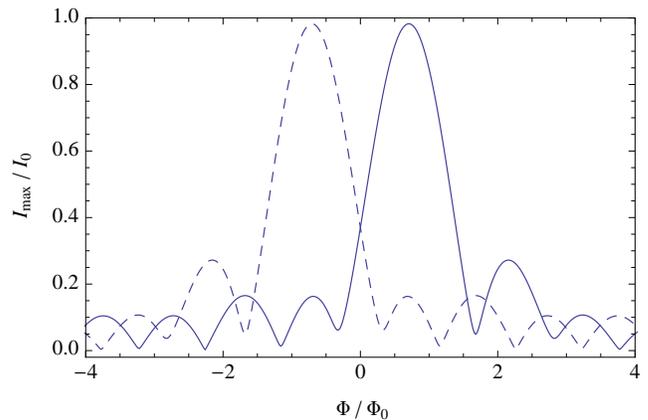}
\end{center}
\caption{\small Shift of the maximum of the critical Josephson current for positive field (solid line) and negative field (dashed line) as observed by Kidwingira \textit{et al} \cite{Maeno2006}.}
\label{Imax_06mRL}
\end{figure}

Starting at zero field, the domain walls shall be essentially randomly configured. Such a situation would lead to a critical current pattern as shown in Fig. \ref{Imax_06m2}. If the magnetic field is increased up to a sufficiently high value, then the domain walls would likely polarize after a certain waiting time. When we decrease now the field to measure the interference pattern, we observe the interference pattern modified by the polarized domain walls. After reaching a sufficiently large negative field value, the domain walls polarize in the opposite way. Therefore, tuning the field back towards positive values we find an altered interference pattern of the critical current. Simulation results taking the two (polarized) domain wall configurations into account are shown in Fig.\ref{Imax_06mRL}. The most striking feature is the shift of the maximum of the critical current. These curves are obtained assuming that changes of domain wall configurations during the field sweep, when the critical currents is measured, can be neglected. 
Reorganizations of the domain wall configurations during the field sweep would most likely lead to discontinuities in the critical current. This kind of behavior is also observed in a set of measurements by Kidwingira et al (Fig.3B in \cite{Maeno2006}) where the field sweeping range is restricted to rather small fields only. Also in this case a hysteretic behavior of critical current (dependence on the field sweep direction) was observed, though less pronounced. 
It is not unlikely that some part of the noise on these data can be interpreted as an effect due to the domain walls.

Turning back to the hysteresis effect with polarized domain wall configurations, we might use the shift of the position of the maximal critical current in order to estimate the density of domain walls. We denote the flux of the maximal current as $ \Phi_{mc} $ which can be determined approximately by the condition
\begin{equation}
\frac{2\pi \Phi_{mc}}{\Phi_0}  \approx - \sum_{i=1}^N \Delta \phi_{i \mu_i} =  -(\phi(L/2) - \phi(-L/2)) = - N \Delta \phi \; ,
\end{equation}
assuming for the last equality a completely polarized domain wall configuration where all domain walls contribute the same phase shift $ \Delta \phi $. 
First, we conclude that the maximum lies at $ \Phi_{mc} > 0 $ for a sweep down from the positive field side (opposite for the opposite sweep direction)  in accordance with experimental findings \cite{Maeno2006}. The number of domain walls intersecting the Josephson junction is then given by
\begin{equation}
N \approx - \frac{ 2 \pi \Phi_{mc}}{\Delta \phi \Phi_0} \; .
\end{equation}
Kidwingira et al find in a measurement of hysteresis effect the magnetic field $ B_{mc} \approx 0.8 G $ (Fig.3A in \cite{Maeno2006}) which  yields a flux $ \Phi_{mc} = d_{eff} L B_{max} \approx 16 \times  G \mu m^2 \approx 0.8 \Phi_0 $. For this estimate of $ \Phi_{mc} $ we took
$ L \approx 100 \mu m $, $ \lambda_{s} = 35 nm $, $ d = 10 nm $ and $ \lambda_{p} = 160 nm $.

Using Eq. (\ref{deltaphi_definition}) with $\nu=-0.6$ we find $\Delta\phi$ in the range of $0.2$ - $0.5$ depending on the orientation of the domain wall with respect to the interface. From this we obtain $ N= 10 $ - $ 25 $ which leads to a mean distance between domain walls of  $ \sim 4 $ - $ 10 \mu m $. 

\begin{figure}[htbp]
\begin{center}
\includegraphics[scale=0.55]{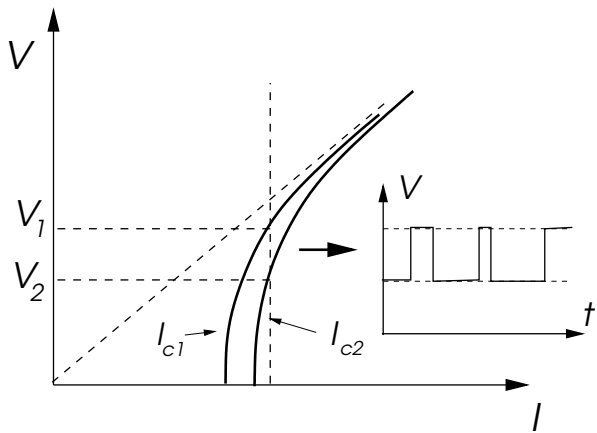}
\end{center}
\caption{\small Schematic I-V characteristic of a Josephson junction. Two values of the critical current are assumed, $ I_{c1} $ and $ I_{c2} ,$ which for the same applied current lead to different values of the voltage on the junction, $ V_1 $ and $ V_2 $, respectively. A time sequence of changing between the two critical current values (by change of domain wall configurations) gives rise to steplike noise on the voltage signal.}
\label{IV}
\end{figure}

Kidwingira et al also report the presence of peculiar noise in the time dependence of the voltage drop of the junction in a constant current slightly above the critical current \cite{Maeno2006}. The question arises whether this feature can also be attributed to the dynamics of domain walls. As we have
seen above, the modification of the domain wall configuration $ \{ \mu_1 , \dots , \mu_N \} $ changes the critical current. The current-voltage characteristics of a Josephson junction for currents $ I $ immediately above the critical current $ I_c $ is very non-linear. The standard textbook form of the current-voltage relation of a Josephson junction gives usually a good approximation of the general behavior
\begin{equation}
V = R \sqrt{I^2 - I_c^2}
\end{equation}
for $ I > I_c $. For a given current $ I $ the time dependence of the critical current, $ I_c(t) = \bar{I}_c + \delta I_c (t) $  leads to
\begin{equation}
\delta V(t) \approx R \bar{I}_c \frac{\delta I_c(t)}{\sqrt{I^2 - \bar{I}_c^2}} \; ,
\end{equation}
which can be large for $ I $ close to $ \bar{I}_c $. The reported noise feature (Fig.3D of \cite{Maeno2006}) suggest that the critical current 
fluctuates between two values, which could correspond to two domain wall configurations (see. Fig.\ref{IV}).

Our theoretical discussion of the interference pattern did not include the faceting of the interface between the $s$-wave superconductor and Sr$_2$RuO$_4$. The effect of faceting is two-fold. First, a faceted surface can lead to additional stable and metastable domain wall configurations as
domain walls can be pinned at surface inhomogeneities. Thus, they add to the number of possible junction states. Second, faceting corresponds to a varying interface normal vector. It is easy to see that the modulation of the inplane normal vector angle $ \theta $ by $ \delta \theta (y)  $ corresponds directly to the Josephson phase variation $ \delta \phi (y) = \delta \theta (y) $ ($ |\delta \theta | \ll \pi $), if the length scale of the faceting is larger than the coherence length of
Sr$_2$RuO$_4$ ($ \xi \sim 80 nm $). As a random phase modulation of this kind with $ | \delta \phi(L/2) - \delta \phi(-L/2) | \leq \overline{|\delta \theta|}  $ does not lead to a significant shift of the maximal critical current in the interference pattern and no hysteretic effect is possible, faceting alone cannot be responsible for the features observed. Together with a spatial variation of the Josephson coupling the variation of the phase due to faceting can contribute to the static random structure of the interference pattern. Fine structures in the interference pattern (variations of the critical current on small field changes) are usually due to variation on long length scales along the  junction and are also known for 
conventional Josephson contacts. 
Note that the effect of faceting in $d$-wave high-temperature superconductors has a stronger impact on the interference pattern, since the phase jumps between 0 and $ \pi $ can lead to phase wandering along the junction such that $ \phi(L/2) - \phi(-L/2) > 2 \pi $ and produces a field shift for the maximal Josephson current, as observed in experiment \cite{MANNHART}.

\section{Long Josephson junction solution}\label{LJJ}

For completeness we address here a further interesting feature connected with a domain wall intersecting the interface. This time, however, we change to limit of $\Lambda_J$ being much shorter than the extension of the junction. In order to describe the behavior of the phase $ \varphi $, we have then to solve the Sine-Gordon equation 
\begin{eqnarray}
\partial_y^2\varphi=\frac{1}{\Lambda_J^2}\sin\varphi+\partial_y^2\phi  \; .
\end{eqnarray}
The solution here has a kink shape which is slightly modified by the last term of the right-hand side.
This term acts like a source term in the vicinity of the domain wall. Without explicitly writing down the solution it is clear that within the length scale of $\Lambda_J$ from the domain wall $ \varphi(y) $  approaches a constant value. As discussed earlier the derivative $ \partial_y \varphi $ corresponds to a local magnetic field,
such that here a well-localized magnetic flux line appears at the position where the domain wall meets the interface. The enclosed magnetic flux can be easily determined by
\begin{eqnarray}
\Phi&=&\int_{-L/2}^{L/2}[A_y(+a,y)-A_y(-a,y)]dy\nonumber\\
&=& \int_{-L/2}^{L/2}\frac{\Phi_0}{2\pi}\left(\partial_y\varphi(y)-\partial_y\phi(y)\right)dy\nonumber\\
&\approx&\frac{\Phi_0}{2\pi}n2\pi-\frac{\Phi_0}{2\pi}\Delta\phi \;,
\end{eqnarray}
where $n$ is a positive integer. We finally see that the minimal possible fluxes is given by
\begin{eqnarray}
\Phi \approx -\frac{\Delta\phi}{2\pi}\Phi_0 , (1- \frac{\Delta\phi}{2\pi}) \Phi_0 \; .
\end{eqnarray}
Since $\frac{\Delta\phi}{2\pi}\leq 1$, we encounter here fractional vortices which are generally a sign of broken time reversal symmetry \cite{Sigrist1999,Sigrist1995}. The detection of such well-localized  fluxes could be used to detect the position of domain walls.

\section{Conclusion}

Motivated by the experiments of Kidwingira et al. on Josephson junctions between Sr$_2$RuO$_4$ and the conventional superconductor Pb \cite{Maeno2006}, we studied effect of domain walls on the Josephson interference effect in a magnetic field assuming that Sr$_2$RuO$_4$ is a chiral $p$-wave superconductor. For this purpose we analyzed the
domain wall structure and showed that its internal phase structure is crucial for the Josephson effect, if domain walls intersect the Josephson junction. The anisotropy of the electronic band structure plays an important role, as it influences the phase shift of the superconducting order parameter between the two types of chiral domains. In addition, it determines the
energetically most favorable orientation of the domain wall. 

The theory presented here is able to explain the most important features reported by Kidwingira et al  \cite{Maeno2006}. (1) The interference pattern in a single junction deviates from the the standard Fraunhofer pattern not only through irregularities, but also shows a distinct asymmetry between positive and negative magnetic fields and  the maximum of the critical current can be shifted away from zero field. This can be attributed to the phase shifts introduced by domain walls intersecting the interface between Sr$_2$RuO$_4$ and the conventional superconductor. (2) Kidwingira et al report strong differences between the interference pattern of different samples. This may be explained by the fact that the samples were prepared in different ways, with possibly also different normal vector directions as well as different degrees of faceting.  (3) Cycling the field continuously covering a positive and negative field range, a hysteretic behavior in the interference pattern appears which can be understood as a field-driven motion of the 
domain walls, whereby the anisotropy of the domain wall energy could play an important role, as our model shows. Allowing for the change of domain wall configurations in time we can also understand the noise effect seen in the current voltage measurements for currents above the critical current. 

The question arises whether it would be possible to test the domain wall scenario by generating a single-domain superconducting phase in Sr$_2$RuO$_4$. This would be most straightforwardly realized by field-cooling. Unfortunately, this procedure might lead to vortex trapping in both superconductors which would jeopardize a clear outcome. Domains naturally arise in the zero-field cooling process, as superconductivity nucleates in the sample with some extent of inhomogeneity, since
the chiral $p$-wave state is rather susceptible to disorder effects. In this way both chiral states can emerge in the sample in different regions and the domain walls appearing between them are eventually pinned in the sample, making it difficult to anneal the sample to a domain wall free phase. It is difficult to estimate a priori the density of domain walls. However, our analysis suggests that domain walls intersecting the Josephson junctions in the experiment may be separated by a few micrometers which is considerably larger than the extension of the domain walls. This estimate may be interesting in the context of recent measurements by Kirtley et al aiming at the observation of spontaneous magnetic flux 
at surfaces of Sr$_2$RuO$_4$, as expected for chiral $p$-wave states \cite{moler}. Since these experiments gave a negative result for the presence of such fluxes, it was speculated that the dense population of chiral domains would cancel out the signal for the scanning SQUID microscope used. Our estimate 
of domain wall density corresponds roughly to the spatial resolution of the SQUID microscope. Thus it remains unclear whether domain walls really could explain the absence of a positive signal from spontaneous flux. Extensive theoretical studies on this issue point to an important puzzle in this context \cite{Ashby}. 

Eventually, we would like to note that this experiment does not decide between even- or odd-parity pairing. While the experimental situation satisfies the selection rules for the coupling between the chiral $p$-wave and an $s$-wave superconductor, the effects observed in the Josephson interference experiments would be similar for a 
chiral $d$-wave superconductor such as the $ k_z (k_x \pm i k_y)$-wave state belonging to $ E_g $-representation of the tetragonal point group $ D_{4h} $, for which, however, additional complications appear due to the nodal gap structure. At the present stage, however, we do not have
any experimental signature in Ref.\cite{Maeno2006}, which would rule out the $E_g $-state.

\section{Acknowledgments}
We are grateful to D.F. Agterberg, Y. Asano, Y. Maeno, M. Matsumoto, T.M. Rice, Y. Tanaka and D. van Harlingen for many enlightening discussions. This work has been financially supported by the Swiss Nationalfonds and the NCCR MaNEP.

\appendix

\section{Analytical approximation}\label{AA}

We start from the free energy functional (\ref{FAxy}) where we neglect the terms in $A_x$. Varying with respect to $\chi$ and $A_y$ gives, respectively,
\begin{eqnarray}
\label{eqchi}
  \partial_y^2\chi &=&\frac{Q}{4\xi_0^2}\sin4\chi-S_+\gamma(\partial_yA_y) \nonumber\\
  &+& C_+[\cos2\chi(\gamma^2A_y^2+(\partial_y\chi)^2)+\sin2\chi(\partial_y^2\chi)]  \;,\\
\label{eqA}
 A_y&=&-\frac{S_+\partial_y\chi}{\gamma(1+C_+\sin2\chi)}  \;.
\end{eqnarray}

Following the argument of Ref.\cite{Sigrist1999} we neglect the contribution of the third term of (\ref{eqchi}) and the $\sin2\chi$ term in the denominator of $A_y$ in (\ref{eqA}).  This leads to the simplified equation for $ \chi(y) $, 
\begin{equation}
 \partial^2_y\chi=\frac{\tilde{Q}}{4}\sin4\chi \;,
\end{equation}
where $\tilde{Q}=Q/[\xi_0^2(1-S_+^2)]$. This has the following kink solution
\begin{equation}
 \chi(y)=\arctan\left(\mathrm{e}^{-\sqrt{\tilde{Q}}y}\right), 
 \label{chi-approx}
\end{equation}
which we use to determine $A_y$. Inserting $\chi$ and $ A_y $ into the free energy functional we derive an analytical form of the variational domain wall energy per unit area,
\begin{eqnarray}
 f(\alpha,\theta)&=&K\eta_0^2\sqrt{\tilde{Q}}\Biggl\{1-\frac{S_+^2}{2}-\frac{C_+\pi}{8}\nonumber\\
 &+&\frac{S_+^2}{C_+}\Biggl[\frac{1}{\sqrt{1-C_+^2}}\arctan\frac{\sqrt{1-C_+^2}}{1+C_+}-\frac{\pi}{4}\Biggr]
\Biggr\} \;.
\end{eqnarray}

\begin{figure}[htbp]
\begin{center}
\includegraphics[scale=0.9]{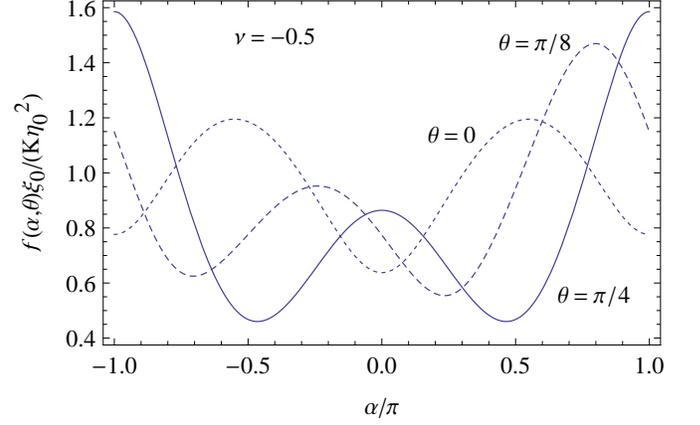}
\end{center}
\caption{\small Domain wall energy (derived from the analytical approximation) for a chosen anisotropy parameter $\nu=-0.5$ as a function of the phase difference $\alpha=\phi_+-\phi_-$ when the angle of the domain wall with respect to the crystal $x$-axis are: $\theta=0,\pi/8$ and $\pi/4$.}
\label{functionala}
\end{figure}

\begin{figure}[htbp]
\begin{center}
\includegraphics[scale=0.9]{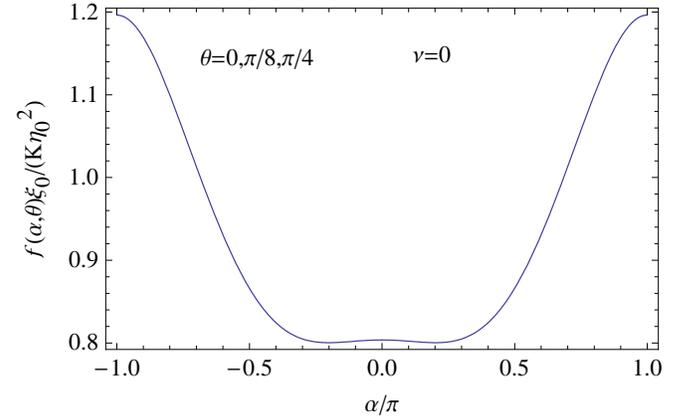}
\end{center}
\caption{\small Domain wall energy derived from the analytical approximation for an isotropic Fermi surface.}
\label{functional0}
\end{figure}

In Fig. \ref{functionala} we show  the domain wall energy as a function of $\alpha$ for the different angles $\theta=0,~\pi/8,~\pi/4$, with the anisotropy parameter chosen $ \nu=-0.5$. We see that those results are qualitatively and even quantitatively close  to the numerical results plotted in Fig.\ref{sola}. 
In Fig. \ref{functional0} we plot the free energy density at $\nu=0$. Here some short-comings of the above approximation becomes obvious, since there a two minima of the energy in contrast to the single one of the numerical result shown in Fig. \ref{sol0}. This discrepancy originates from  neglecting the vector potential $A_x$ and the corresponding terms in the free energy.

\section{Numeral solutions for $\nu=-0.6$}\label{sol_eta_phi}

\begin{figure}[H]
\begin{center}
\includegraphics[scale=1.35]{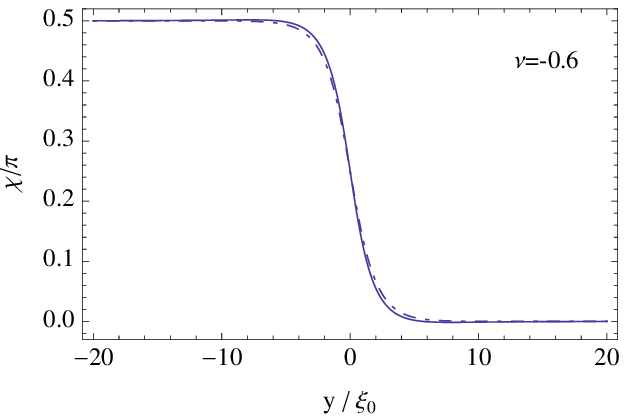}
\end{center}
\caption{\small Spacial dependence of $\chi(y)$ through a domain wall, calculated numerically (solid line) and with the analytical approximation (dot-dashed line) for $\nu=-0.6$.}
\label{ChiAN}
\end{figure}

\begin{figure}[H]
\begin{center}
\includegraphics[scale=1.35]{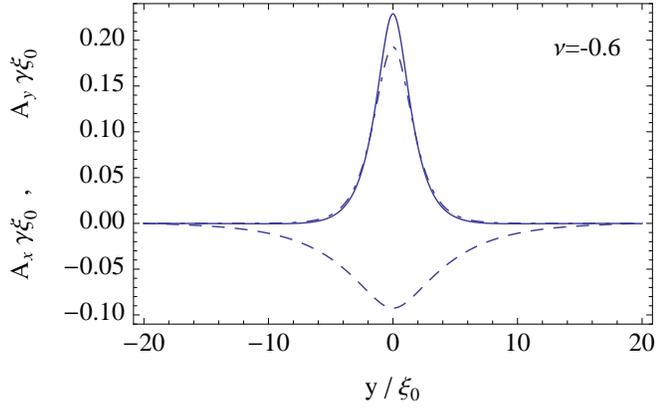}
\end{center}
\caption{\small Spacial dependence of the vector potential through a domain wall. Numerical solutions of $A_y$ (solid line) and $A_x$ (dashed line). Analytical solution of $A_y$ (dot-dashed line).}
\label{A_xy06m}
\end{figure}

We show here the numerical solutions of the structure of the domain wall in the most stable state for $\nu=-0.6$ and compare with the analytical approximation. We plot in Fig. \ref{ChiAN} the function $\chi(y)$ where the solid line is  the numerical solution and the dot-dashed line the approximative solution in Eq. (\ref{chi-approx}). The deviation is rather small. In Fig. \ref{A_xy06m} we show the two components of the vector potential, the solid line for the numerical and the dot-dashed line for the approximate solution. 
Here the discrepancy is larger.

\begin{figure}[H]
\begin{center}
\includegraphics[scale=1.35]{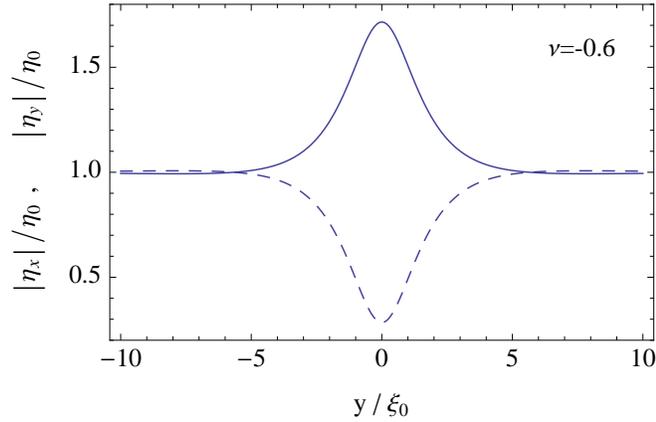}
\end{center}
\caption{\small Spacial dependence of the modulus of the order parameter through a domain wall : $\vert\eta_x\vert$ (dashed line), $\vert\eta_y\vert$ (solid line).}
\label{eta06m}
\end{figure}

\begin{figure}[H]
\begin{center}
\includegraphics[scale=1.35]{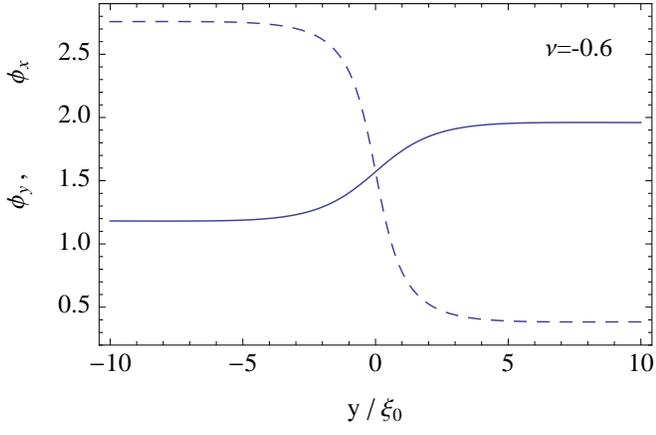}
\end{center}
\caption{\small Spacial dependence of the phase of the order parameter through a domain wall : $\phi_x$ (dashed line), $\phi_y$ (solid line).}
\label{phi06m}
\end{figure}

Fig. \ref{eta06m} and \ref{phi06m} depict the modulus of both order parameter components, $ |\eta_x| $ and $ | \eta_y | $ and the phase of the order parameter passing through the domain wall (we choose $\phi_{\pm}=\pm\alpha/2$). Obviously, the two phase differences follow the relation $\Delta\phi_x = \alpha-\pi$ and $\Delta\phi_y=\alpha$. 

\end{document}